\documentclass[amssymb,aps,prd]{revtex4-2}
\usepackage{graphicx}
\usepackage{subfig}
\usepackage{mathtools}
\usepackage{url}
\usepackage[normalem]{ulem}
\usepackage{caption}

\usepackage[colorlinks=true, linkcolor=blue, citecolor=blue, urlcolor=blue]{hyperref}

\newcommand{\be}{\begin{equation}}
\newcommand{\ee}{\end{equation}}
\newcommand{\bea}{\begin{eqnarray}}
\newcommand{\eea}{\end{eqnarray}}
\newcommand{\ba}[1]{\begin{array}{#1}}
	\newcommand{\ea}{\end{array}}
\newcommand{\nn}{\nonumber}

\newcommand{\Om}{\Omega}

\newcommand{\del}{\partial}
\newcommand{\orcid}[1]{\href{https://orcid.org/#1}{\includegraphics[width=8pt]
{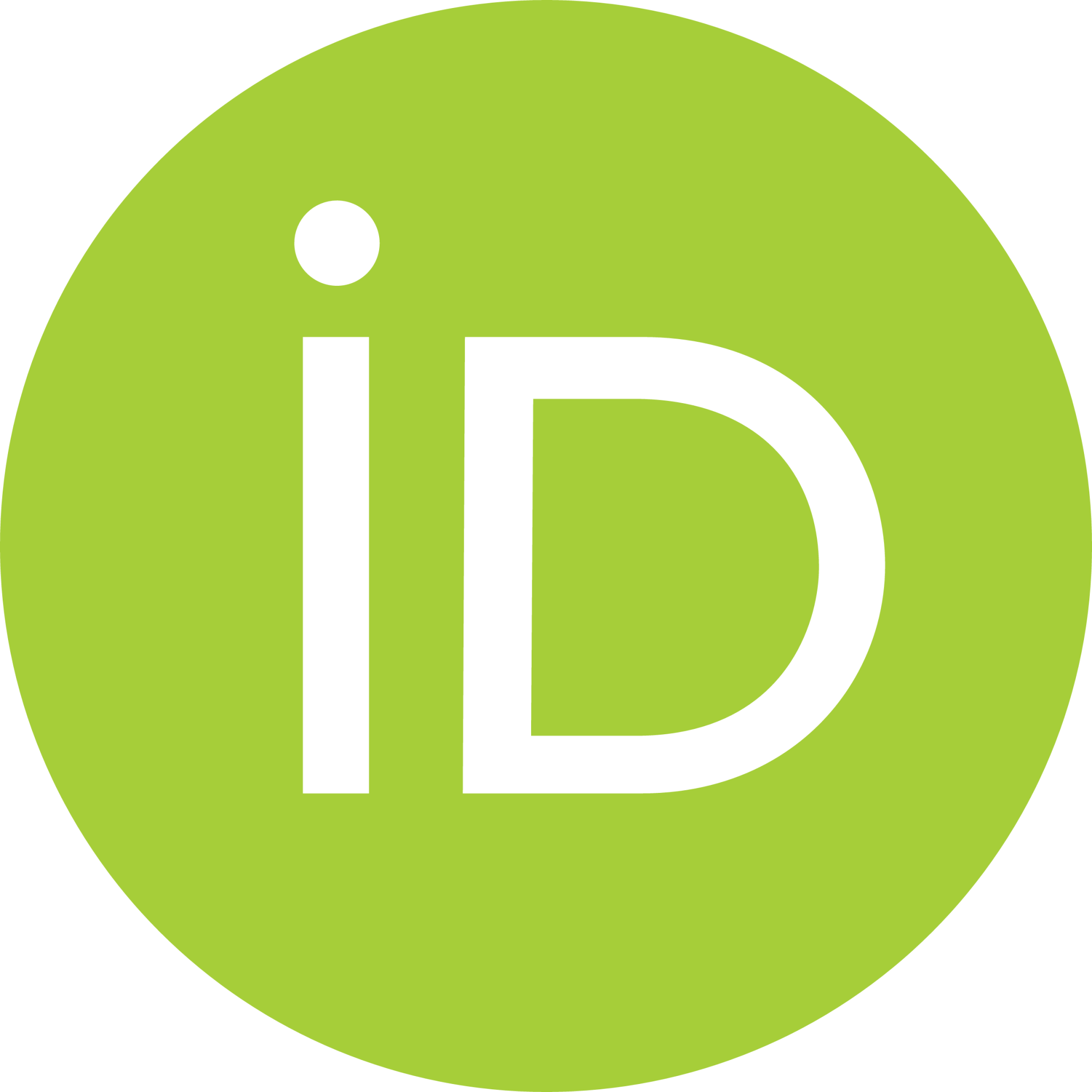}}}

\begin{document}
 \title{Effect of Coriolis Force on Diffusion of D Meson}
\author{Ashutosh Dwibedi\orcid{0009-0004-1568-2806}$^1$}
\author{Nandita Padhan\orcid{0009-0008-6857-2650}$^2$}
\author{Dani Rose J Marattukalam\orcid{0009-0006-8204-8148}$^1$}
\author{Arghya Chatterjee\orcid{0000-0002-2895-6085}$^1$}
\author{Sudipan De$^3$}
\author{Sabyasachi Ghosh\orcid{0000-0003-1212-824X}$^1$}
\thanks{Corresponding author: sabyaphy@gmail.com}

 \affiliation{$^{1}$Department of Physics, Indian Institute of Technology Bhilai, Kutelabhata, Durg, 491002, Chhattisgarh, India}
\affiliation{$^{2}$Department of Physics, National Institute of Technology Durgapur, Durgapur, 713209, West Bengal, India}
\affiliation{$^{3}$Department of Physics, Dinabandhu Mahavidyalaya, Bongaon, North 24 Parganas, 743235, West Bengal, India}
%

%
\begin{abstract}
We have attempted to calculate and estimate the spatial diffusion
coefficients of D meson through rotating hadron resonance gas, which can be produced in the late stage of peripheral heavy ion collisions. Employing the framework of kinetic theory in relaxation time approximation, and using Einstein’s diffusion relation, one can  express the spatial diffusion coefficients of D meson as a ratio of its conductivity to its susceptibility. Here, we have tuned D meson relaxation time from the knowledge of earlier works on its spatial diffusion estimations, and then we have extended the framework for the finite rotation picture of hadronic matter, where only the effect of Coriolis force is considered. Our study also revealed the anisotropic nature of diffusion in the presence of rotation with future possibilities of phenomenological signature.

     \end{abstract}
\maketitle
\section{Introduction}

Quark-Gluon Plasma (QGP), a state of QCD matter, is expected to form in ultra-relativistic heavy-ion collisions (HICs)~\cite{Shuryak:2004cy,Wong:1995jf}. Hard partons and heavy quarks (HQs) are profusely produced at Relativistic Heavy Ion Collider (RHIC) and Large Hadron Collider (LHC) energies in initial hard scattering processes. They are sensitive probes of the medium formed in the collision as they are produced at the early stages of the collision and witness the entire evolution of the QGP~\cite{Gossiaux_2008,Gossiaux_2009,Gossiauxto_2009,Nahrgang_2015,DAS2015260,PhysRevC.96.044905,Chandra:2024ron,Pooja:2023gqt,Prakash:2023wbs}. HQs may lose energy by collisions as well as by gluon bremsstrahlung while propagating through the medium~\cite{Thoma:1991jum,Mustafa:1997pm,Gossiauxe_2009,Gossiaux_2010,Kumar:2021goi,Ruggieri:2022kxv,Prakash:2021lwt}. The energy loss of the HQs inside the medium can be quantified by measuring the transverse momentum suppression ($R_{AA}$)~\cite{ALICE:2021rxa}, which may be considered as an indirect measurement of the drag and diffusion coefficient of charm quark in QGP~\cite{Prino:2016cni,Rapp:2018qla,Das:2010tj}. For a comprehensive review of different transport models of HQ and hadronization mechanisms, one can see the Refs.~\cite{Cao_2019,zhao2023hadroni}. HQ transport in the pre-equilibrium phase of the QGP has also been studied in Refs.~\cite{Boguslavski:2020tqz,Pooja:2022ojj,Boguslavski:2023fdm,Pooja:2024rnn,Backfried:2024rub}.
An extremely strong magnetic field (of the order of $10^{18}$ to $10^{19}$ Gauss~\cite{Tuchin:2013ie}) is expected to produce in peripheral  heavy-ion collision~\cite{Skokov:2009qp,Bzdak:2011yy}. A strong magnetic field may influence the relativistic fluids, and thus flow~\cite{Das:2016cwd}, jet quenching coefficient~\cite{Banerjee:2021sjm}, heavy quark/ meson diffusion coefficients~\cite{Satapathy:2022xdw},  etc. may also be affected. 
The diffusion phenomenology of the heavy quarks and meson at a finite magnetic field was discussed in Ref.~\cite{Satapathy:2022xdw}.
In off-central HICs, along with the creation of huge magnetic fields, a large orbital angular momentum (OAM) can also be transferred from the initial colliding nuclei to the formed medium~\cite{Liang:2004ph,Becattini:2007sr}. 
In this respect, the medium formed in off-central collisions at the RHIC can be considered to be a rotating system, possessing a significant OAM ranging up to $10^7\hbar$~\cite{STAR:2017ckg,Liang:2004ph,Becattini:2007sr}. This initial OAM subsequently manifests itself as local vorticity first in the quark fluid and later in the hadronic fluid. The vorticity can result in various effects such as spin polarization~\cite{BECATTINI201332}, the chiral vortical effect (CVE) \cite{PhysRevLett.103.191601}, etc. 
The STAR Collaboration measured the global spin polarization of $\Lambda$ and $\Bar{\Lambda}$ particles in Au + Au collisions over a range of collision energies ($\sqrt{s_{NN}}=$ 7.7-200 GeV), revealing a decreasing trend with collision energies~\cite{STAR:2017ckg}. A recent study with better statistics at $\sqrt{s_{NN}}=$ 200 GeV found that polarization depends on event-by-event charge asymmetry. This suggests that the axial current induced by the initial magnetic field might contribute to global polarization~\cite{STAR:2018gyt}. Furthermore, spin alignment has been observed in vector mesons, with recent measurements at the RHIC and LHC enhancing our understanding of spin phenomena in heavy ion collisions~\cite{ALICE:2019aid,STAR:2022fan,ALICE:2022dyy}. Moreover, the presence of Coriolis force in a rotating medium can also lead to anisotropic diffusion coefficients of HQs and mesons as previously studied in the presence of Lorentz force because of magnetic fields~\cite{Satapathy:2022xdw} in the laboratory frame. In the present work, we will focus on the diffusion phenomenology of heavy mesons due to the presence of Coriolis force in the rotating medium.

There is a notable connection between rotational effects and magnetic fields, both of which can arise in off-central collisions. The Coriolis force, arising from rotation, and the Lorentz force, generated in the presence of magnetic fields, exhibit striking similarities in their effects on moving particles Refs.~\cite{J_Sivardiere_1983, Johnson2000-px,Sakurai1980}, though the former acts in the rotating frame while the latter operates in the inertial (laboratory) frame. The presence of magnetic fields induces anisotropy in the transport coefficients and this anisotropy in the context of viscosities and conductivities of the produced nuclear matter in HICs has been investigated in Refs.~\cite{Dey:2019axu,Dash:2020vxk,Dey:2020awu,Ghosh:2019ubc,Dey:2019vkn,Kalikotay:2020snc,Dey:2021fbo,Satapathy:2021cjp,Das:2019wjg,Das:2019ppb,Dey:2019axu,Chatterjee:2019nld,Hattori:2016cnt,Hattori:2016lqx,Satapathy:2021wex}. 
Apart from these studies of transport coefficients of QGP and hadronic gases in the presence of magnetic fields, the effect of magnetic fields in the dynamics of HQs inside the QGP and the hadronic system has also been studied. Initial studies on the dynamics of $J/\psi$ mesons have used holographic QCD to explore the influence of magnetic fields on charmonium~\cite{Dudal:2014jfa}. 
Simplified holographic QCD models have also been employed to examine the transport properties of $J/\psi$ and heavy quarks, showing that spatial diffusion is split into longitudinal and transverse components based on the direction of the magnetic field~\cite{Dudal:2018rki, Satapathy:2022xdw}.
 Due to the similarity in the mathematical expression of Coriolis force and Lorentz force, one can expect similar type of transport coefficients in a medium viewed from the rotating frame and in a medium subjected to a magnetic field in the inertial (laboratory) frame.
In connection to this the Refs.~\cite{Aung:2023pjf, Dwibedi:2023akm,Padhan:2024edf,Dwibedi:2025boz,Dwibedi:2025bdd} have explored the the mathematical similarity between the Coriolis force and Lorentz force to calculate the anisotropic transport coefficients of QGP and hadronic matter. 
Specifically, the anisotropic nature of shear viscosity and electrical conductivity in the presence of the Coriolis force was observed in Refs.~\cite{Aung:2023pjf, Dwibedi:2023akm} in a non-relativistic framework. 
The effect of Coriolis force in the electrical conductivity and shear viscosity of a hadron gas was also studied in a relativistic framework by employing the Hadron Resonance Gas (HRG) model in Ref.~\cite{Padhan:2024edf,Dwibedi:2025bdd} and in Nambu--Jona-Lasinio model in Ref.~\cite{Dwibedi:2025boz}. Aside from the transport phenomena of rotating QGP and hadronic medium, the HQs and mesons transport can be significantly affected inside a rotating QGP or hadronic medium. In particular, the diffusion coefficients of heavy mesons can also exhibit a similar structure in the rest frame of a rotating medium as it exhibits in the presence of magnetic fields~\cite{Dudal:2018rki,Satapathy:2022xdw} in the inertial (laboratory) frame. Nevertheless, the dynamics of heavy mesons inside a rotating medium have not been addressed thoroughly in the literature. 
 %
%
Earlier studies on diffusion calculations in the absence of magnetic filed or medium rotation primarily focused on heavy-quark diffusion with in the QGP medium; a detailed list of such studies can be found in Refs.~\cite{Prino:2016cni,Rapp:2018qla,Das:2010tj,Rahaman:2020tha}. In contrast, the diffusion of heavy mesons and baryons within the hadronic matter was generally overlooked earlier (particularly before 2010), due to assumptions about its negligible impact. It was until Refs.~\cite{PhysRevC.90.054909,PhysRevD.84.011503,Das:2011vba,Ghosh:2014oia} highlighted non-negligible contributions of heavy-flavored meson or baryon diffusion in the hadronic matter. For a recent and comprehensive review on the transport properties of open heavy-flavor hadrons in hot hadronic matter, readers are referred to Ref.~\cite{Das:2024vac}.
In this work, we extend this investigation by incorporating the effects of rotation on the diffusion of heavy mesons within the hadronic medium. We particularly focus on the diffusion of  D meson, as it is the lightest meson carrying heavy flavor and this serves as an ideal probe for such studies. As an initial step, we calculate the anisotropic spatial diffusion coefficients, leaving the actual evolution of the D meson distribution in the rotating frame with the aid of Fokker-Planck or Langevin equation as a future work. To fulfill our goal, we first write the relativistic Boltzmann transport equation (BTE) for D meson distribution in a rotating frame of reference by adhering to the relaxation time approximation (RTA). The generalization of BTE to the rotating frame of references has been made with the aid of connection coefficients, which in turn can be calculated from the space-dependent rotating frame metric. The relaxation time of the D meson with the background hadronic gas has been calculated by the popular HRG model. The HRG is a well-established model for describing the hadronic phase of matter produced in relativistic heavy ion collisions. In this framework, the system can be effectively treated as a multi-species gas consisting of various particles such as protons, neutrons, and pions along with the numerous unstable resonant states documented by the Particle Data Group~\cite{ParticleDataGroup:2008zun}. The HRG model has been widely used to explore a wide range of phenomena, including HIC thermodynamics~\cite{Karsch:2003zq,Vovchenko:2014pka,Vovchenko:2019pjl,Pradhan:2023rvf} and fluctuations of conserved charges~\cite{Begun:2006jf,Nahrgang:2014fza,HotQCD:2012fhj,Bhattacharyya:2013oya,Chatterjee:2016mve,Vovchenko:2017ayq}. The HRG model has proven valuable in estimating various transport coefficients that govern the system's response to external forces ~\cite{Greif:2016skc,Prakash:1993bt,Gorenstein:2007mw,Noronha-Hostler:2012ycm,Tiwari:2011km,Pradhan:2022gbm,Ghosh:2014qba,Dwibedi:2024mff,Wiranata:2014kva,Noronha-Hostler:2008kkf,Kadam:2014cua,Rose:2017bjz}. In this investigation, we use an ideal HRG model for the estimation of D meson diffusion in rotating HRG medium.

The article is arranged as follows. In Sec.~(\ref{NRDif}), we recapitulate the conventional framework of diffusion and give a layout for calculating the spatial diffusion coefficients of D meson from the BTE. Afterward, in Sec.~(\ref{Ddiff}), we first develop the covariant BTE in the rotating frame by illustrating the different kinds of forces that affect the meson transport. Secondly, we calculate the spatial diffusion tensor of the meson with the help of BTE in RTA. Then, in Sec.~(\ref{sec:HRGmodel}), we briefly describe the HRG model, which we use to determine the D meson relaxation time by assuming hard sphere scattering interactions. In Sec.~(\ref{Sec:Results_Discussion}), we display the numerical estimations of the conductivity and diffusion of the D meson and quantitatively discuss the anisotropy produced because of the rotation. Ultimately, we summarize our findings in Sec.~(\ref{Sec:Summary}). In the end, an appendix discusses the detailed derivation of heavy meson conductivity from the relativistic Boltzmann equation.

\section{\textbf{Framework and Model descriptions}}\label{MD}
\subsection{Non-Rotating Framework of Diffusion}\label{NRDif}
Let us first briefly review the traditional methods of describing the heavy meson or heavy quark phenomenology before going into the details of our methodology. Present work only focuses on heavy meson diffusion through hadronic matter, although the primarily practiced heavy quark diffusion through QGP also shares the same mathematical structure. This existing framework is done by not considering any rotational effects of the medium, so we are calling it a non-rotating framework (name of this subsection). Most of the literatures~\cite{RAPPB,DONG201997,HE2023104020} in the topics of  heavy meson or heavy quark diffusion determine the evolution of the heavy meson or heavy quark momenta (or, distribution function) through the help of Langevin equation (or, equivalent, Fokker-Planck equation). The momentum drag and diffusion coefficients serve as input to the evolution equations and determine the momenta (or, distribution) at a later time provided initial momentum $\vec{p} ~(t=0)$ (or, momentum space distribution $f(\vec{r},~\vec{p},~t=0)$) of the meson or quark is known. These two approaches are equivalent~\cite{Risken1996,zwanzig2001nonequilibrium,Debbasch1997}.
		In the following we will first describe the Langevin dynamics of D meson and then the equivalent Fokker-Planck dynamics in a concise manner. The Langevin equation for a D meson traveling inside a static thermal background is given by~\cite{Svetitsky:1987gq,Moore:2004tg},
		\bea
		&& \frac{d p^{i}}{dt}=-\eta^{ij}_{D}~p^{j}+\xi^{i}(t)~,\label{lan1}\\
		&& \frac{d x^{i}}{dt}=\frac{p^{i}}{E}~ {\rm with}~E=\sqrt{\vec{p}^{2}+m^{2}_{D}} ~,\label{lan2}
		\eea
		where $x^{i}$, $p^{i}$, $E$, and $m_{D}$ are, respectively, the position, momentum, energy, and mass of the D meson. The first term in the RHS of Eq.~(\ref{lan1}) is known as the drag term, and the second term is a stochastic force term that arises because of the random thermal kicks D meson receives from the thermal background. The stochastic/fluctuating force is specified by its ensemble average $\langle\xi^{i}\rangle=0$ and correlations $\langle\xi^{i}(t)~\xi^{j}(t^{\prime})\rangle=2 ~B^{ij} ~\delta(t-t^{\prime})$. The $B^{ij}$ and $\eta_{D}^{ij}$ are identified as the momentum diffusion and momentum drag coefficients, respectively, in the literature~\cite{Svetitsky:1987gq,Moore:2004tg}. For a radially expanding system likely to be produced in HIC, one needs to appropriately boost the solutions of Eqs.~(\ref{lan1}) and ~(\ref{lan2}) by the fluid velocity of the expanding media to get the position and momentum variables in the lab frame~\cite{He:2013zua}. After determining heavy meson position and momentum in the lab frame, one can calculate various experimental observables like $R_{AA}$, $v_{2}$, etc, of the heavy meson under consideration. In the Fokker-Planck approach~\cite{vanHees:2004gq,Dask2009,Das:2010tj,Mazumder:2011nj} one writes a partial differential equation for distribution function $f(\vec{r},~\vec{p},~t)$ as,
		\bea
		\frac{p^{\mu}}{E}\frac{\del f}{\del x^{\mu}}=\frac{\del}{\del p^{i}}\bigg[A_{i} f +\frac{\del}{\del p^{j}} B_{ij}f\bigg]~,\label{fp1}
		\eea
		where $A^{i}\equiv \eta^{ij}_{D}~p^{j}$.
		The interpretation of the distribution function is probabilistic, i.e., $f(\vec{r},~\vec{p},~t)~ d^{3}\vec{p} ~d^{3}\vec{r}$ gives us probability to find the meson around the phase space point $(\vec{r},~\vec{p})$ at time $t$. Eqs.~(\ref{lan1}) and ~(\ref{lan2}) are equivalent to Eq.~(\ref{fp1}); therefore, one may solve either of them to get the heavy flavor observables. We notice from the Langevin or the Fokker-Planck equations that the momentum drag $\eta_{D}^{ij}$ and diffusion $B^{ij}$ coefficients serve as valuable inputs for the evolution of heavy mesons inside the medium. The coupling of heavy mesons with the background QCD matter can also summarized by providing the spatial diffusion coefficients, which gives a measure of asymptotic mean squared displacement traversed by the meson initially placed at a point (say origin)~\cite{Capellino2022}. For the case of an isotropic fluid, one obtains a simplified set of heavy meson transport coefficients in $\vec{p}\xrightarrow{}{}0$ limit. In this limit, the spatial diffusion coefficient $D^{ij}_{s}=D_{s}~\delta^{ij}$, momentum drag coefficient $\eta^{ij}_{D}=\delta^{ij}\gamma_{D}$ and momentum diffusion coefficient $B_{ij}=D~\delta^{ij}$ are connected via the following relations: $D=m_{D}\gamma_{D}T$, $D_{s}=\frac{T}{m_{D}\gamma_{D}}$ and $D_{s}=\frac{T^{2}}{D}$~\cite{Svetitsky:1987gq,Moore:2004tg,Capellino2022}. Therefore, one can get the spatial diffusion coefficient knowing the momentum diffusion coefficient and vice versa.
In the following, we outline the method to calculate the spatial diffusion coefficients $D_{s}^{ij}$ of D meson diffusing inside the light hadronic matter. This should be perceived as a first step to qualitatively understand the diffusion of the D meson in a rotating hadron gas. The continuity equation for the D meson current can be written as,
		\bea
		\del_{\mu} J^{\mu}=0, \text{ where } J^{\mu}=\int f ~p^{\mu} \frac{d^{3}\vec{p}}{p_{0}}~,\label{conti}
		\eea
		where $p_{0}\equiv E$ is the energy of the D meson. Eq.~(\ref{conti}) guarantees the number conservation of D meson and is valid in the full non-equilibrium scenario. Nevertheless, to calculate the spatial diffusion coefficients associated with the diffusion, one can assume a slight perturbation $\delta f$ over the equilibrium distribution $f_{0}$ of the D meson, which can be accomplished by creating a spatial gradient in D meson chemical potential $\mu_{D}$ around $\mu_{D}=0 $~\cite{Petreczky:2005nh,Satapathy:2022xdw}. The total four current $J^{\mu}$ in this picture can be broken into two parts: $J^{\mu}\equiv J_{0}^{\mu}+\delta J^{\mu}$, where the first part $J_{0}^{\mu}\equiv(n_{0},J_{0}^{i}=0)$ comes from the equilibrium distribution $f_{0}$ and the second part $\delta J^{\mu}\equiv(\delta n=0,\delta J^{i})$ comes from the out-of-equilibrium distribution $\delta f$. Using the definition provided in Eq.~(\ref{conti}), the microscopic expression for the out-of-equilibrium current density $\delta J^i$ of the  D meson diffusing through a hadronic background can be written as 
		\bea
		\delta J^{i} &=& \int\frac{d^3\vec{p}}{(2\pi)^3}\frac{p^i}{p_0}~\delta f~.\label{micon}
		\eea
		Moreover, in the Navier-Stokes limit, one can write the macroscopic expression of the out-of-equilibrium current density as,
		\bea
		\delta J^{i}=-\sigma^{ij} ~\nabla^{j}\mu_{D}\equiv -D^{ij}_{s}~\chi ~\nabla^{j}\mu_{D}= -D^{ij}_{s}~\nabla^{j}n~,\label{macon}
		\eea
		where the D meson density $n=\int f_{0} ~\frac{d^{3}\vec{p}}{(2\pi)^{3}}$ and susceptibility $\chi=\frac{\del n}{\del \mu_{D}}$. The Eq.~(\ref{micon}) provides the kinetic theory definition of the D meson current, which can be evaluated by determining the out-of-equilibrium distribution $\delta f$ by solving the BTE. On the other hand, Eq.~(\ref{macon}) is reminiscent of Ohm's law with the driving force as the electric field had replaced by the negative gradient of $\mu_{D}$. The role of the spatial diffusion coefficients will be clear by substituting the expression of $J^{i}$ in Eq.~(\ref{conti}) which implies the Fick's law of diffusion, $\frac{\del n}{\del t}=\nabla^{i}\nabla^{j}D_{s}^{ij}n$. After evaluating $J^{i}$ from Eq.~(\ref{micon}) with the help of BTE, one can obtain the  D meson conductivity tensor $\sigma^{ij}$ by comparing Eq.~(\ref{micon}) with Eq.~(\ref{macon}). Then, the diffusion coefficients of D meson can be obtained by taking the ratio of its conductivity tensor with susceptibility. Their standard expressions are~\cite{Satapathy:2022xdw}, $\sigma=  \frac{1}{3T}\int \frac{d^{3}p}{(2\pi)^3}\tau_c\times \frac{p^2}{E^{2}}f_0(1+f_0)$, $\chi= \frac{1}{T}\int\frac{d^3p}{(2\pi)^3}f_0(1 + f_0)$ and $D_{s}=\left[\int d^{3}p~\tau_c\times \frac{p^2}{E^{2}}f_0(1+f_0)\right]/\left[3\int d^3p~ f_0(1 + f_0)\right]$, where $\tau_{c}$ is the relaxation time of D meson in the hadronic medium.

After this quick recapitulation of existing non-rotating framework of diffusion, we will now proceed to the next subsection for the explicit calculation of D meson spatial diffusion coefficients $D_{s}^{ij}$ from the BTE in the rotating frame.

\subsection{\textbf{Rotating Framework of Diffusion}}\label{Ddiff}	
In this section, our final goal will be to write down the covariant BTE in the rotating frame and evaluate the spatial diffusion coefficients for the D meson. Before moving towards our final goal, we will briefly describe the physical picture and the mathematical tools needed in the procedure; readers can get the detailed mathematical framework in Ref.~\cite{Padhan:2024edf}.

In the previous works related to transport in the rotating nuclear medium, the Refs.~\cite{Aung:2023pjf, Dwibedi:2023akm} have explored the structure of shear viscosity and electrical conductivity of a rotating QGP using Non-relativistic BTE. This calculation for the rotating nuclear matter has been extended recently to the relativistic realm in Ref.~\cite{Padhan:2024edf}, where the covariant BTE is used to obtain the anisotropic conductivities for hadronic gas employing the popular HRG model. All these models have a common physical picture in which one explicitly incorporates the rotational background of the medium expected in off-central HIC in the kinetic description. Subsequently, one writes down a BTE in the rotating frame to calculate the transport properties of the QGP and the hadronic gas. Here, in contrast to the aim of the Refs.~\cite{Aung:2023pjf,Dwibedi:2023akm,Padhan:2024edf}, we will be concerned with the diffusion of open charmed mesons (D meson) through the rotating hadronic matter.  In order to address this diffusion phenomenon, the mathematical framework of Ref.~\cite{Padhan:2024edf} can be borrowed with following important changes: the equation of motion (EOM) in the rotating frame will be that of the  D meson which diffuse in the background light hadrons and the covariant BTE will be set up for the distribution function of D meson to determine the diffusion coefficients.

Let us consider the motion of the D meson from the perspective of the rotating medium which rotates around the $z-$axis with angular velocity $\vec{\Omega}$ relative to the lab-fixed inertial frame. Any point on the rotating medium at a distance $r$ from the axis rotates with the speed $\Omega~ (\sqrt{x^{2}+y^{2}})\equiv \Omega r$, which should be less than speed of light $c\equiv 1$ to respect the causality~\cite{Chernodub:2017ref}. Therefore, we should restrict the spatial region transverse to the rotation axis within the causal cylinder given by, $\Omega r<1$.
 The coordinate transformation from inertial coordinates ${\bf x}\equiv(t, x, y, z)$ to rotating coordinates ${\bf x}^{\prime}\equiv(t^{\prime}, x^{\prime}, y^{\prime}, z^{\prime})$ ~\cite{Chernodub:2016kxh,Chernodub:2017ref,Chernodub:2020qah,FUJIMOTO2021136184}:
\\
\begin{equation}
	{\bf x}^{\prime} = \mathbf{R}(\Omega t) ~{\bf x},\label{A1}
\end{equation}
is essential to describe the EOM of D meson in a rotating frame, where the rotation matrix $\mathbf{R}(\Omega t)$ for transforming coordinates from the inertial frame to the rotating frame is given by:

\bea
&&\mathbf{R}(\Omega t) = 
\begin{pmatrix}
	1 & 0 & 0 & 0 \\
	0 & \cos(\Omega t) & \sin(\Omega t) & 0 \\
	0 & -\sin(\Omega t) & \cos(\Omega t) & 0 \\
	0 & 0 & 0 & 1
\end{pmatrix}\label{A2}.
\eea
Using the transformation law provided in Eq.~(\ref{A1}) and (\ref{A2}) one can obtain the squared length element $ds^{2}$, metric tensor $g_{\mu\nu}$ and connection coefficients $\Gamma_{\mu \lambda}^{\alpha}$ as follows~\cite{Padhan:2024edf}:
\bea
&&ds^{2}= g_{\mu\nu} dx^{\prime\mu}dx^{\prime\nu}=dt^{\prime^{2}}(1-\Om^2x^{\prime^2}-\Om^2y^{\prime^2})+2\Om y^{\prime}dt^{\prime}dx^{\prime}-2\Om x^{\prime} dt^{\prime}dy^{\prime}-dx^{\prime^2}-dy^{\prime^2}-dz^{\prime^2},\nn\\
&& g_{\mu\nu}=
\begin{pmatrix}
	1-\Om^2x^{\prime^2}-\Om^2y^{\prime^2}  & \Om y^{\prime} & -\Om x^{\prime} & 0\\
	\Om y^{\prime}                           &      -1        &         0       & 0\\
	-\Om x^{\prime}                           &       0        &        -1       & 0 \\
	0                                        &       0        &         0       & -1 
	\label{A3}
\end{pmatrix},\\
&&\Gamma_{\mu \lambda}^{\alpha}=\frac{1}{2}g^{\alpha \nu}\left(\frac{\del g_{\nu \mu}}{\del x^{\lambda}} +\frac{\del g_{\lambda \nu}}{\del x^{\mu}} - \frac{\del g_{\mu \lambda}}{\del x^{\nu}}\right).\label{A4}
\eea 
Now, we are ready to write down the EOM for the  D meson, which will eventually be required to establish the BTE for the  D meson diffusion. The EOM for the  D meson in the rotating frame is given by \cite{Cercignani200210,misner2017gravitation,schutz2009first}:
\begin{equation}
	\frac{d p^{\alpha}}{d\tau} + \frac{1}{m_{D}} p^{\mu} p^{\lambda} \Gamma_{\mu \lambda}^{\alpha} =0, \label{A5}
\end{equation}
where $p^{\alpha}$ and $\tau$ are the four-momentum and proper time, respectively. The non-zero connection coefficients in the present case can be obtained by resorting to Eq.~(\ref{A3}) and (\ref{A4}) as follows \cite{Kapusta:2019sad}: $\Gamma_{00}^{1}=-\Om^{2}x, \Gamma_{00}^{2}=-\Om^{2}y, \Gamma_{20}^{1}=\Gamma_{02}^{1}=-\Om, \Gamma_{10}^{2}=\Gamma_{01}^{2}=\Om$. Let us recast Eq.~(\ref{A5}) with the substitution of non-zero connection coefficients in order to observe the resemblance between the EOM of D meson supplied by Eq.~(\ref{A5}) and the classical nonrelativistic EOM in the rotating frame \cite{goldstein2011classical,kleppner2014introduction},

\begin{equation}
	\frac{d \vec{p}}{dt} =  2 \gamma_{v} m_{D} (\vec{v} \times \vec{\Omega})+....~, \label{A6}
\end{equation}
where the four-momentum is given by, $p^{\alpha} = (\gamma_v m_{D}, \gamma_v m_{D} \vec{v}) = (\gamma_v m_{D}, \vec{p})$ with the Loretz factor $\gamma_v = \frac{dt}{d\tau}$. A quick glance at Eq.~(\ref{A6}) suggests that the first term in the RHS of Eq.~(\ref{A6}) is the relativistic version of the Coriolis force. In the present paper, we have ignored the other possible pseudo forces.
 Now, we are equipped with all the necessary tools to write down the BTE for D meson diffusing under the rotating hadronic background. The covariant BTE in the co-rotating frame can be written as:
\bea
&&p^{\mu} \frac{\partial f}{\partial x^{\mu}} - m_{D} \frac{dp^{\alpha}}{d\tau}\frac{\partial f}{\partial p^{\alpha}}=C[f] \nn\\
\implies &&p^{\mu} \frac{\partial f}{\partial x^{\mu}} - \Gamma_{\mu \lambda}^{\alpha} p^{\mu} p^{\lambda}\frac{\partial f}{\partial p^{\alpha}}= - (u^{\alpha} p_{\alpha}) \frac{\delta f}{\tau_{c}}, \label{A7}
\eea
where we used Eq.~(\ref{A5}) and approximated the collision kernel $C[f]$ by the RTA i.e., $C[f]\thickapprox  - (u^{\alpha} p_{\alpha}) \frac{\delta f}{\tau_{c}}$ to get the last equality. The $\tau_{c}$ that appears in the collision kernel approximated by RTA is the average time of collision between  D meson and the HRG system. The total distribution $f$ of the D meson can be written as $f=f_{0}+\delta f$, where $f_{0}$ is given by, 
\begin{equation}
	f_0 = \frac{1}{e^{(p^{\alpha}u_{\alpha} - \mu_{D}) / T} - 1},\label{A8}
\end{equation}
where $p^{\alpha}$ is the four momenta, $u^{\alpha}=(\frac{1}{\sqrt{g_{00}}}, 0)$ is the static fluid four velocity. 
Eq.~(\ref{A7}) can be solved to find out $\delta f$ and subsequently the conductivities of  D meson. In solving Eq.~(\ref{A7}) we consider only the Coriolis force. Here, we write the final expression of conductivities with the detailed derivations provided in Appendix~(\ref{sec:ECB}). In the case where there is no rotation of the medium, the conductivity tensor may be written as $\sigma_{ij} = \sigma \delta_{ij}$; however, an anisotropic conductivity tensor can be generated in the presence of rotation from non-relativistic~\cite{Dwibedi:2023akm} to relativistic~\cite{Padhan:2024edf} calculation whose form is given by, 

\be 
\sigma^{ij}=\sigma^0 ~\delta^{i j} +\sigma^1 ~\epsilon^{ijk} \omega^k + \sigma^2 ~\omega^i \omega^j,\label{ancon}
\ee 
where $\epsilon^{ijk}$ is the Levi-Civita symbol and $\omega^i$ is a unit vector along the angular velocity $\vec{\Omega}$, i.e., $\vec{\Omega}\equiv \Omega \hat{\omega}$, which is now considered in an arbitrary direction but one can go to the special case $\vec{\Omega}=\Omega \hat{k}$ for understanding the phenomenological picture. The nonzero components of the anisotropic conductivity tensor $\sigma^{ij}$ are related to each other as,
\bea 
&&{\rm Perpendicular/Transverse~ component:}~ \sigma^{xx}=\sigma^{yy}=\sigma^{0}\equiv\sigma^{\perp},
\nn\\
&&{\rm Hall~ component:}~\sigma^{xy} = -\sigma^{yx}=\sigma^{1}\equiv\sigma^{\times},
\nn\\
&&{\rm Parallel/Longitudinal~ component:}~\sigma^{zz}=\sigma^{0}+\sigma^{2}\equiv\sigma^{\parallel}.\label{ancon1}
\eea 
By using RTA-based kinetic theory formalism~\cite{Satapathy:2022xdw,Padhan:2024edf}, one can get this multicomponents conductivity of D meson (see Appendix~(\ref{sec:ECB})). The parallel (or, the  D meson conductivity in the absence of $\Omega$), perpendicular, and the Hall conductivity of the  D meson are respectively given by,
\bea
&&\sigma^{zz} =\sigma^{||}=  \frac{1}{3T}\int \frac{d^{3}p}{(2\pi)^3}\tau_c\times \frac{p^2}{E^{2}}f_0(1+f_0),\label{s_HRG}\\
&&\sigma^{xx} = \sigma^{yy}=\sigma^{\perp} =\frac{1}{3T}\int \frac{d^{3}p}{(2\pi)^3}\frac{\tau_c}{1+\big(\frac{\tau_c}{\tau_{\Omega}}\big)^2}\times \frac{p^2}{E^{2}}f_0(1+f_0), (\tau_{\Omega}\equiv 1/2\Omega)\label{sO_HRG}\\
&&\sigma^{xy} = -\sigma^{yx}=\sigma^{\times} = \frac{1}{3T}\int \frac{d^{3}p}{(2\pi)^3}\frac{\tau_c\big(\frac{\tau_c}{\tau_{\Omega}}\big)}{1+\big(\frac{\tau_c}{\tau_{\Omega}}\big)^2}\times \frac{p^2}{E^{2}}f_0(1+f_0),\label{sOH_HRG}
\eea
where $f_{0}=1/(e^{E/T}-1)$ is the Bose-Einstein distribution function for D meson.
 In presence angular velocity $\vec{\Om}$ the spatial diffusion coefficients become anisotropic and take a $3\times 3$ matrix structure provided by,
\bea
D_{s}^{ij} = \frac{\sigma^{ij}}{\chi}~,
\label{D_1}
\eea 
where $\sigma^{ij}$ can be obtained from the formula given in Eqs.~(\ref{s_HRG}) to (\ref{sOH_HRG})  and the susceptibility $\chi$, which is defined as:
\be
\chi =\frac{\del n}{\del \mu_{D}}= \frac{1}{T}\int\frac{d^3p}{(2\pi)^3}f_0(1 + f_0).
\label{chi_2}
\ee
Using Eqs.~(\ref{s_HRG}) to (\ref{sOH_HRG}) in Eq.~(\ref{D_1}), we get the expressions for parallel, perpendicular, and Hall diffusion coefficients as,
\bea
&&D^{||}_{s}= \frac{\sigma^{||}}{\chi} =  \frac{\frac{1}{3T}\int \frac{d^{3}p}{(2\pi)^3}\tau_c\times \frac{p^2}{E^{2}}f_0(1+f_0)}{\frac{1}{T}\int\frac{d^3p}{(2\pi)^3}f_0(1 + f_0)},\label{D_paral}\\
&&D^{\perp}_{s} = \frac{\sigma^{\perp}}{\chi} =
\frac{\frac{1}{3T}\int \frac{d^{3}p}{(2\pi)^3}\tau^{\perp}_{c}\times \frac{p^2}{E^{2}}f_0(1+f_0)}{\frac{1}{T}\int\frac{d^3p}{(2\pi)^3}f_0(1 + f_0)},\label{D_perp}\\
&&D^{\times}_{s} = \frac{\sigma^{\times}}{\chi} = \frac{\frac{1}{3T}\int \frac{d^{3}p}{(2\pi)^3}\tau^{\times}_{c}\times \frac{p^2}{E^{2}}f_{0}(1+f_0)}{\frac{1}{T}\int\frac{d^3p}{(2\pi)^3}f_0(1 + f_0)},\label{D_hall}
\eea
where $\tau^{\perp}_{c}\equiv\frac{\tau_c}{1+\big(\frac{\tau_c}{\tau_{\Omega}}\big)^2}$ and $\tau^{\times}_{c}\equiv\frac{\tau_c\big(\frac{\tau_c}{\tau_{\Omega}}\big)}{1+\big(\frac{\tau_c}{\tau_{\Omega}}\big)^2}$ are respectively the effective relaxation time of D meson in perpendicular and Hall directions. The readers can notice that due to finite $\tau_{\Omega}$ originated from finite Coriolis force, we will get anisotropy in D meson conductivity ($\sigma^{||}\neq\sigma^{\perp}$) and diffusion coefficients ($D^{||}_{s}\neq D^{\perp}_{s}$) and also their non-vanishing Hall components ($\sigma^{\times}\neq 0~, D^{\times}_{s}\neq 0 $). In the limit $\vec{\Omega}\rightarrow 0$, we get an isotropic conductivity $\sigma^{||}=\sigma^{\perp}(\Omega\rightarrow 0)=\sigma$ (say) and diffusion $D^{||}_{s}=D^{\perp}_{s}(\Omega\rightarrow 0)=D_{s}$ (say). For the complete determination of diffusion coefficients provided in Eqs.~(\ref{D_paral}) to (\ref{D_hall}), we need to specify the relaxation time $\tau_{c}$ of the D meson, which measures the interaction of the D meson with the hadronic gas. For this purpose, we will model the hadronic gas with the HRG model and interactions of D mesons with HRG with a hard sphere scattering model, which is addressed in the next section.

\subsection{\textbf{HRG model and relaxation time of D Meson}}
\label{sec:HRGmodel}

The HRG model is a widely accepted framework for characterizing the hadronic phase of matter resulting from relativistic heavy ion collisions~\cite{KARSCH2011136,GARG2013691,PhysRevC.92.054901,PhysRevC.101.035205,PhysRevC.94.014905,PhysRevC.90.024915}. This model offers a statistical depiction of hadrons and resonances using the grand canonical ensemble approach. At sufficiently high temperatures, the kinetic energy predominates over inter-hadronic interactions, causing hadrons and resonances to behave like an ideal gas of non-interacting particles. We have used the Ideal Hadron Resonance Gas (IHRG) model for this work. In the IHRG model, the partition function accounts for all relevant degrees of freedom associated with the system. Using S-matrix calculations, it has been shown that in the presence of narrow resonances, the thermodynamics of the interacting gas of hadrons can be approximated by an ideal gas of hadrons and their resonances~\cite{Dashen:1969ep,Dashen:1974jw}. Here, it comprises point-like hadrons up to mass 2.6 GeV as listed in Ref.~\cite{ParticleDataGroup:2008zun}. 
The thermal system produced in heavy-ion collider experiments bears a resemblance to the grand canonical ensemble.
The thermodynamic variables like pressure ($P$), particle number density ($n$), energy density($\epsilon$), entropy density(s), etc, of the produced thermal system can be expressed in terms of the partition function (Z).

In the present work, only the total number density of the HRG system will be directly used for the calculation of D meson relaxation time. According to the standard grand canonical ensemble framework, one can obtain the number density from its partition function, and we can write it as a summation of meson and baryon contribution:

\be
n_{HRG}=\sum_{B}g_B\int\frac{d^3p}{(2\pi)^3}\frac{1}{e^{E/T}+1} +\sum_{M} g_M\int\frac{d^3p}{(2\pi)^3}\frac{1}{e^{E/T}-1}~. 
\label{n_H}
\ee
When D meson diffuses through the HRG matter, it will face the HRG matter density $n_{\rm HRG}$ during collisions. So, one can define D meson relaxation time in HRG matter as:
\be
\tau_c=1/(n_{\rm HRG} v_{\rm av}\pi a^2),
\label{tau_c}
\ee 
where,
\be 
v_{\rm av}=\int \frac{d^3p}{(2\pi)^3}\frac{p}{E} f_0\Big/\int \frac{d^3p}{(2\pi)^3} f_0\label{av_vel}
\ee
is the average velocity of D meson and $\pi a^2$ is considered to be the hard sphere cross section for the  D meson, having scattering length $a$.

\section{Results and Discussion}
\label{Sec:Results_Discussion}
\begin{figure*}[b]
    \centering
    \includegraphics[width=0.46 \textwidth]{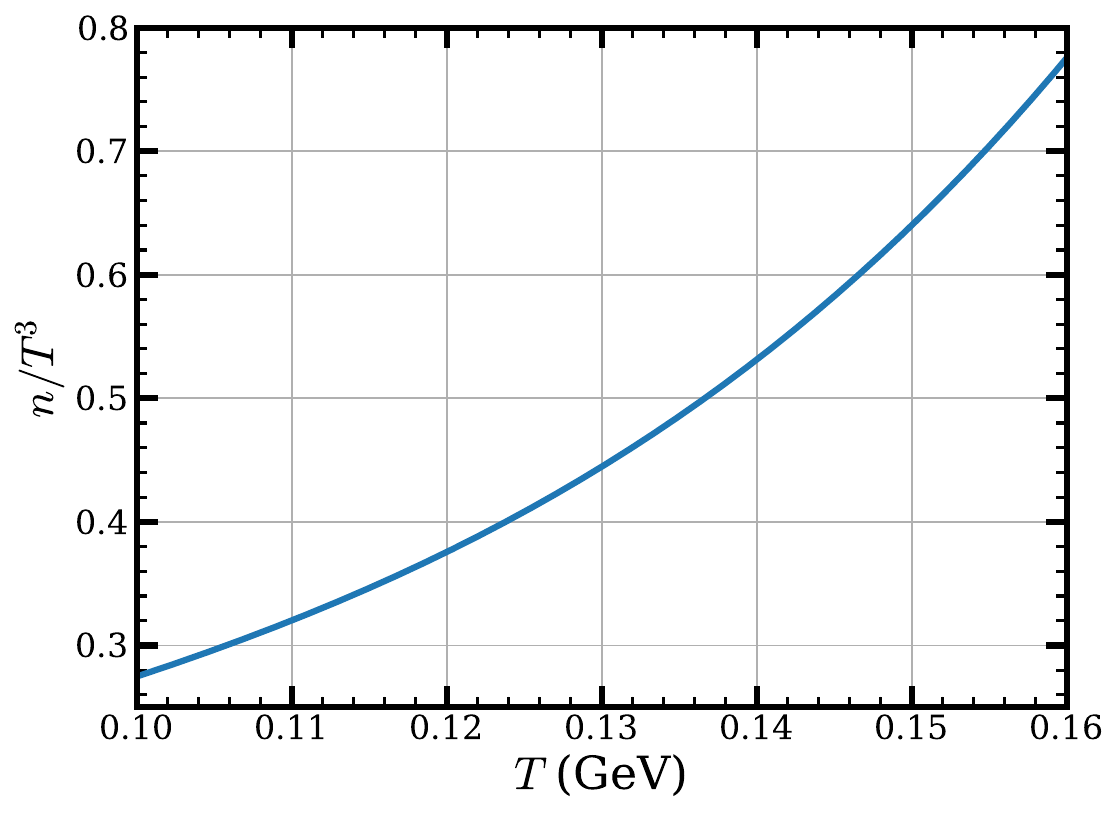}
    \includegraphics[width=0.46 \textwidth]{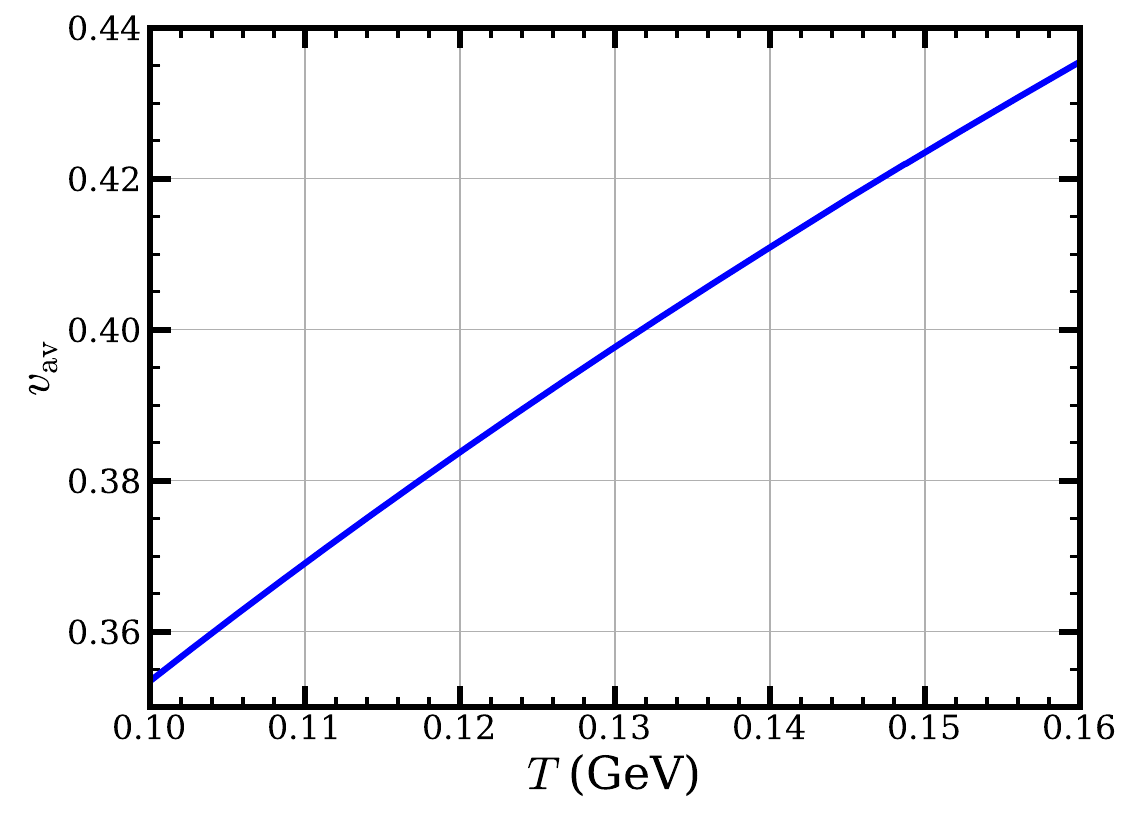}
    \captionsetup{justification=centering, singlelinecheck=off}
    \caption{(Color online) Left: Number density as a function of temperature. Right: Average velocity of D meson as a function of temperature.}
    \label{figure1}
\end{figure*}

In this section, we have numerically studied the influence of rotation on the spatial diffusion of D mesons in hadronic matter using the IHRG model, which comprises all the non-interacting hadrons and their resonances up to mass 2.6 GeV as listed in Ref.~\cite{ParticleDataGroup:2008zun}. 
To understand the rotational effect, we have investigated the results of D meson diffusion coefficients in a rotating medium of hadrons and compared them with those when the medium was not rotating. We have used the RTA to calculate the diffusion coefficient, which is defined as the ratio of conductivity to susceptibility, referring to Eq.~(\ref{D_1}). 
The relaxation time of the D meson depends on its velocity and the system's number density as described in Eq.~(\ref{tau_c}).

\begin{figure}
	\centering
	\includegraphics[width=0.45\textwidth]{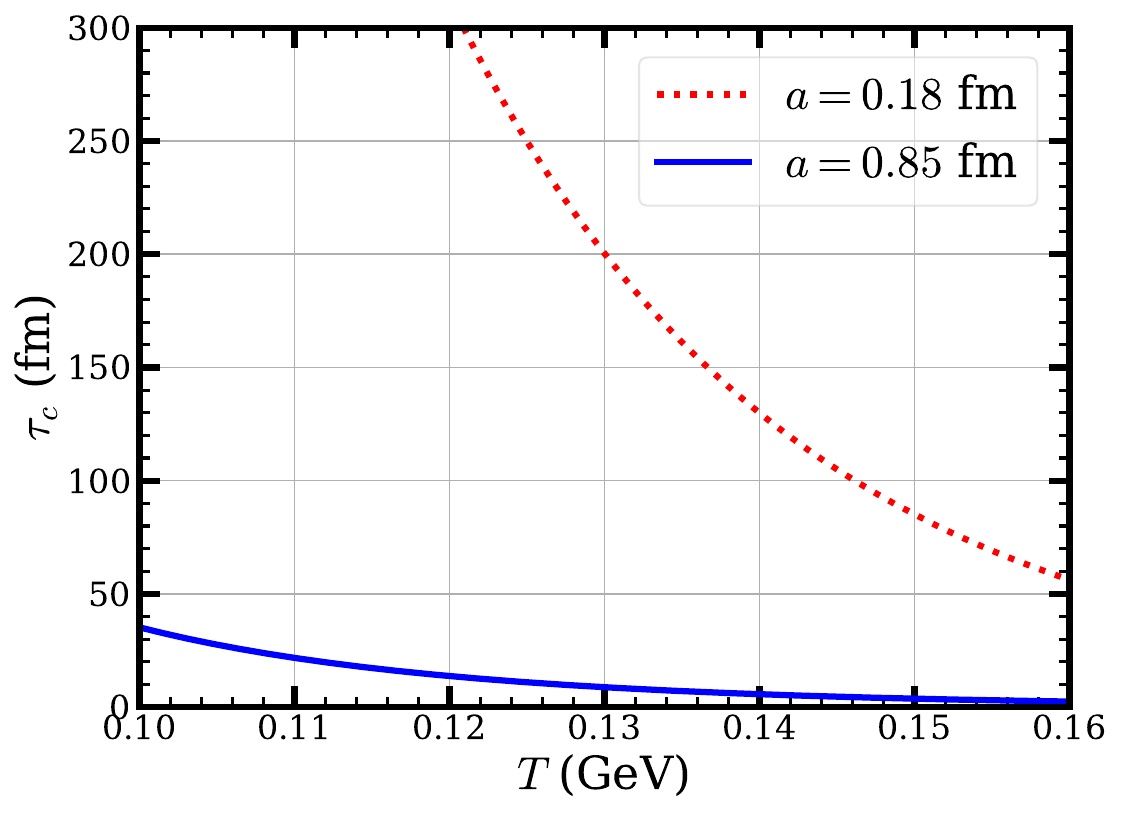}
	\includegraphics[width=0.48\textwidth]{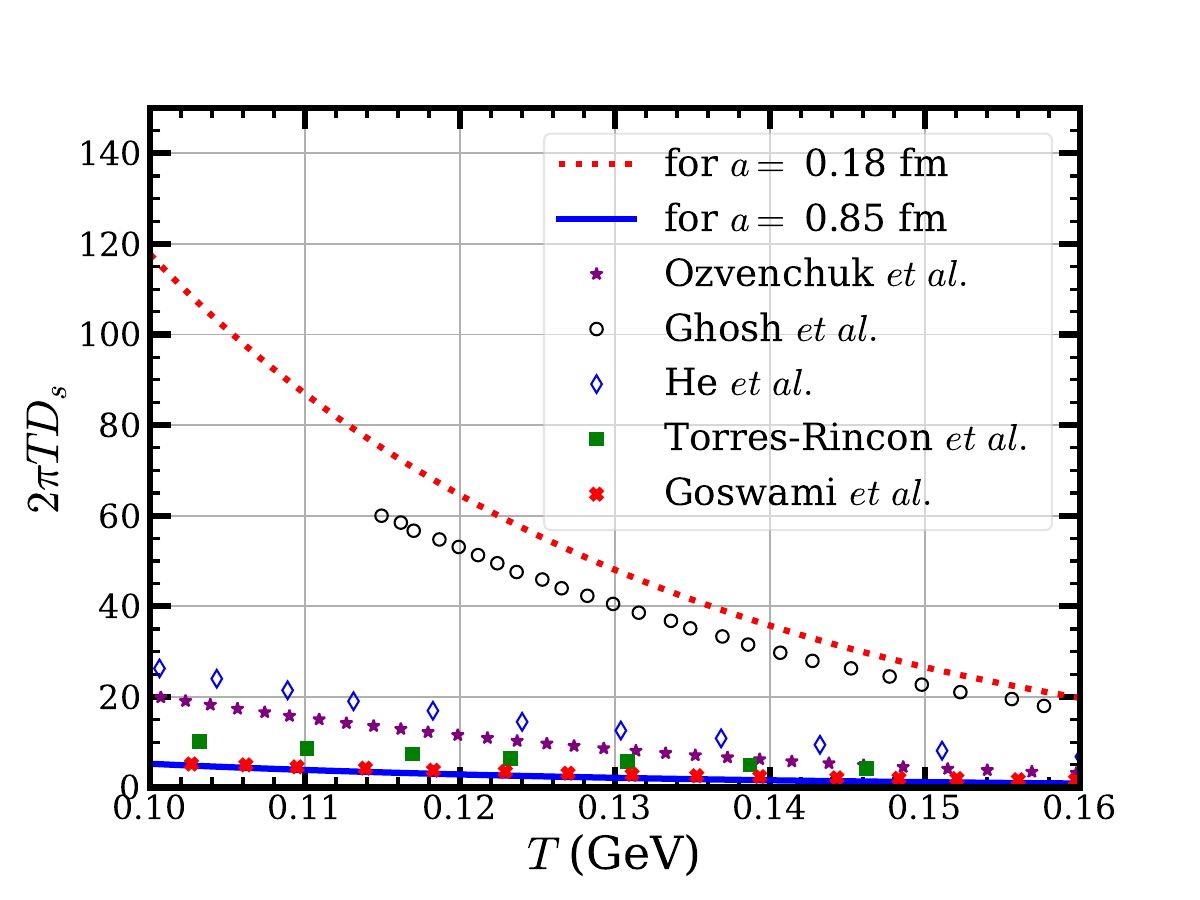} 
	\captionsetup{justification=centering, singlelinecheck=off}
	\caption{(Color online) Left: Relaxation time of D meson as a function of temperature. Right: spatial diffusion coefficient ($D_{s}$) for D meson as a function of temperature and comparing the result with Ozvenchuk et al.~\cite{PhysRevC.90.054909}, Ghosh et al.~\cite{PhysRevD.84.011503}, He et al.~\cite{He:2011yi}, Torres-Rincon et al.~\cite{Torres-Rincon:2021yga}, and Goswami et al.~\cite{Goswami:2023hdl}. }
	\label{figure2}
\end{figure}

The temperature-dependent number density profile of the HRG system and the temperature variation of the D meson velocity are presented in the left and right panels of Fig.~(\ref{figure1}), respectively. Our numerical estimation suggests that the average velocity of  D mesons within the hadron gas medium increases with $T$ in the hadronic temperature domain (up to 0.16 GeV), ranging from 0.35 to 0.44. As observed in the left panel of Fig.~(\ref{figure1}), the reader can guess a $T^b$-dependence (with $b>3$) of the actual number density of HRG, and hence, a $T^{-b}$-dependence of relaxation time can be expected as shown in the left panel of Fig.~(\ref{figure2}). The scattering length $a$ is an important parameter for estimating $\tau_c$. Here, we have taken $a=0.18$ and $0.85$ fm to analyze the variation on the order of magnitude of $\tau_c$.

In the left panel of Fig.~(\ref{figure2}), we have presented the temperature-dependent relaxation time ($\tau_{c}$) depicted within the hadronic region using the hard sphere scattering model. As described in Eq.~(\ref{tau_c}) of Sec.~(\ref{sec:HRGmodel}), $\tau_{c}$ is inversely proportional to number density, velocity, and scattering length. Using two different scattering lengths: $a$=0.18 fm and $a$=0.85 fm (the reason for choosing these two specific values of $a$ will be clear later), we have shown that the relaxation time decreases as the temperature increases as a result of increased number density. At a particular temperature, the relaxation time falls as the scattering length increases, suggesting stronger particle interactions.

In the right panel of Fig.~(\ref{figure2}) we have illustrated the temperature dependence of the scaled diffusion coefficient ($2\pi T D_{s}$). The figure shows that $2\pi T D_{s}$ decreases as the temperature increases, which gives a similar nature to the earlier spatial diffusion data from Ghosh et al.~\cite{PhysRevD.84.011503} (open black circles), Ozvenchuk et al.~\cite{PhysRevC.90.054909} (solid purple stars),  He et al.\cite{He:2011yi} (open blue diamond), Torres-Rincon et al.\cite{Torres-Rincon:2021yga} (green square), and Goswami et al.\cite{Goswami:2023hdl} (red cross). To cover these earlier estimations, we have taken two different values of the scattering length, $a = 0.18$ fm and $0.85$ fm, for the computation of $D^{||}_{s}\equiv D_{s}$ from Eq.~(\ref{D_paral}). The magnitude of $2\pi T D_{s}$ decreases with temperature for a given scattering length, this is a result of decreased relaxation time which stems from the enhanced hadronic number density with temperature. Furthermore, the figure also demonstrates that as the scattering length increases, the value of $2\pi T D_{s}$ decreases further, indicating that a lower diffusion coefficient resulting from stronger particle interactions. After tuning the relaxation time (by tuning $a$) to cover the earlier estimation of the diffusion coefficient in the absence of the rotation, we will now proceed to show the variation of perpendicular and Hall conductivity and diffusion coefficients as a function of $\Omega$ and $T$.

\begin{figure}[t]
	\centering
	\includegraphics[width=0.45\textwidth]{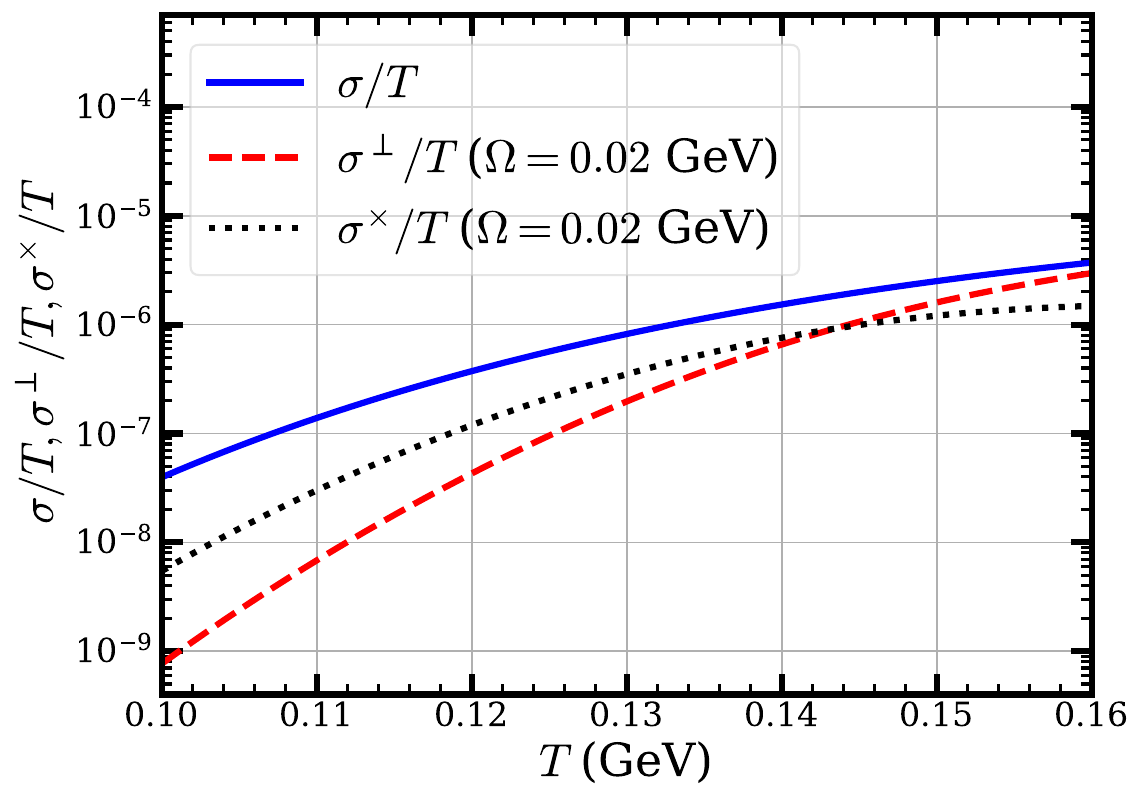}
	\includegraphics[width=0.45\textwidth]{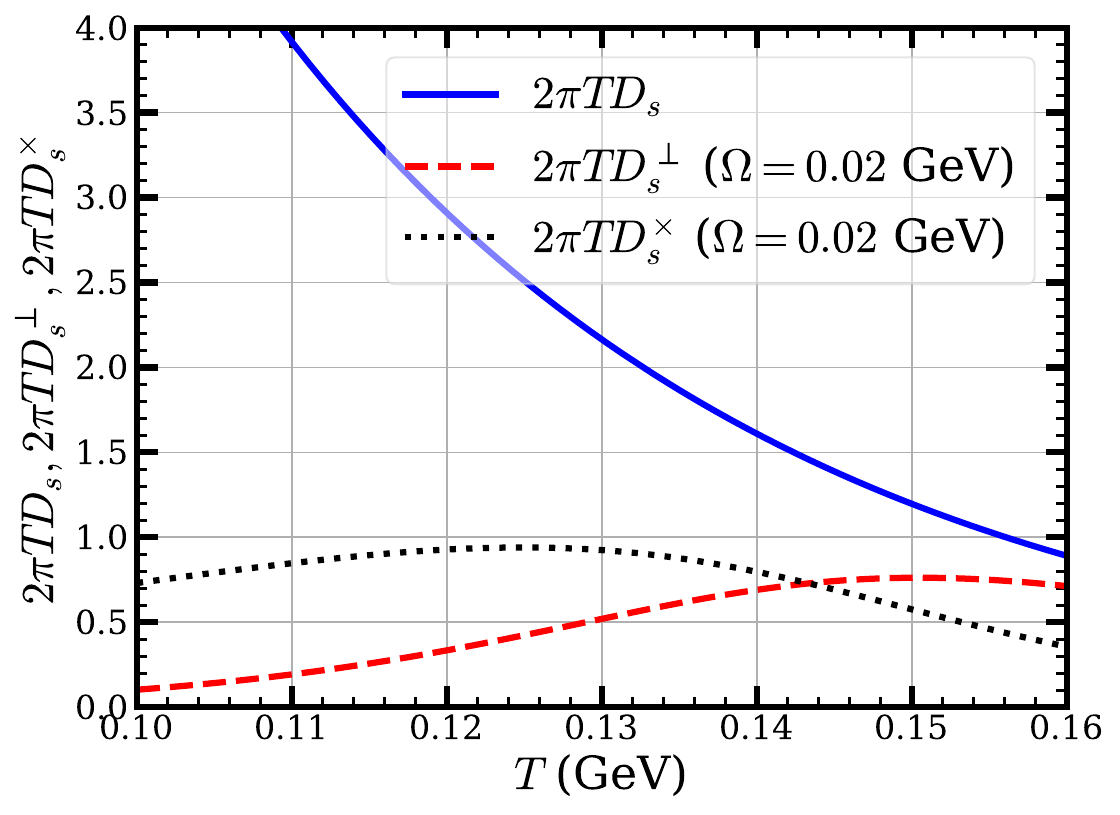}
	\captionsetup{justification=centering, singlelinecheck=off}
	\caption{(Color online) Left: Parallel, perpendicular and Hall conductivity of D meson ($\sigma / T$, $\sigma^{\perp} / T$, $\sigma^{\times} / T$) as functions of temperature. Right:  Parallel, perpendicular, and Hall spatial diffusion coefficients ($2\pi T D_{s}$, $2\pi T D^{\perp}_{s}$, $2\pi T D^{\times}_{s}$) as functions of temperature for D meson.}
	\label{figure3}
\end{figure}

We have presented the temperature dependence of the scaled conductivity component of D mesons in the left panel of Fig.~(\ref{figure3}): the parallel ($\sigma$ / $T$, represented by the blue solid line), perpendicular ($\sigma^{\perp}$ / $T$, represented by the red dashed line) and Hall ($\sigma^{\times}$ / $T$, represented by the black dotted line) conductivities.  
These components are evaluated at a constant hard-sphere scattering length of $a = 0.85$ fm and a rotational time scale of $\tau_{\Omega} = 5$ fm (corresponding to $\Omega$ = 0.02 GeV). The parallel conductivity remains unchanged under rotation, while the perpendicular and Hall conductivities are affected by the rotational conditions. These calculations are performed using the framework discussed in Eq. (\ref{s_HRG}-~\ref{sOH_HRG}), in Sec.~(\ref{Ddiff}). 
It can be noted that in the temperature region, $T \in (0.1, 0.15)$ GeV,  all three components of conductivity increase. In the right panel of Fig.~\ref{figure3}, the temperature dependence of the diffusion components of D meson, i.e. parallel ($2\pi T D_{s}$), perpendicular ($2\pi T D^{\perp}_{s}$) and Hall ($2\pi T D^{\times}_{s}$) are shown at a constant value of $a = 0.85$ fm and $\tau_{\Omega} = 5$ fm. 
Referring to Eq.~(\ref{D_paral}), (\ref{D_perp}), and (\ref{D_hall}) — it becomes evident that rotation does not affect the susceptibility ($\chi$); only the conductivity is modified. Here, we find that $2\pi T D_{s}$ approach close to $2\pi T D^{\perp}_{s}$ as the temperature increases. To understand the nature of the curves in Fig.~(\ref{figure3}), it is convenient to express the conductivity as well as the spatial diffusion components in the following way: $\sigma^{||,\perp,\times}=\tau_{c}^{||,\perp,\times}~(P.S.)_{\sigma}$ and $D_{s}^{||,\perp,\times}=\tau_{c}^{||,\perp,\times}~(P.S.)_{D}$ with $\tau_{c}^{||}\equiv \tau_{c}, (P.S.)_{\sigma}= \frac{1}{3T}\int \frac{d^{3}p}{(2\pi)^3}\times \frac{p^2}{E^{2}}f_0(1+f_0) $, and  $(P.S.)_{D}=\frac{(P.S.)_{\sigma}}{\chi}$.
		In general, the behavior of conductivity or diffusion components against temperature is determined by two competing factors--effective relaxation times ($\tau_{c}^{||,\perp,\times}$) and the thermodynamic phase-space ($P.S.$) part. The relaxation time, $\tau_{c}^{||}=\tau_{c}$ decreases with temperature, whereas the variation of the effective relaxation times $\tau_{c}^{\perp,\times}$ has some attractive features. It is interesting to note the behavior of $\tau_{c}^{\perp,\times}$ in the two extreme limits:
		\begin{itemize}
			\item $\tau_{c}\gg\tau_{\Omega}$, i.e., when the relaxation time of D meson is very large than the rotational time scale,
			\item $\tau_{c}\ll\tau_{\Omega}$, i.e., when the rotational time scale dominates the relaxation time.
		\end{itemize}
		The first scenario is achieved at lower temperatures (cf. Fig.~(\ref{figure2})) and one has $\tau_{c}^{\perp}\approx\frac{\tau_{\Omega}}{\tau_{c}}\tau_{\Omega}$ and $\tau_{c}^{\times}\approx\tau_{\Omega}$. The second scenario is prevalent at higher temperatures, and one has $\tau_{c}^{\perp}\approx\tau_{c}$ and $\tau_{c}^{\times}\approx \tau_{c}^{2}/\tau_{\Omega}$. Coming to the phase space part, $(P.S.)_{\sigma}$ increases rapidly with $T$ leading to the sharp rise of $\sigma^{||,\perp,\times}$ one encounters in the left panel of Fig.~(\ref{figure2}). On the other hand, the phase space part $(P.S.)_{D}=\frac{(P.S.)_{\sigma}}{\chi}$ of the spatial diffusion coefficients increase at a much slower rate compared to $(P.S.)_{\sigma}$ because the rapid increase of $\chi$ cancels the sharp increase in $(P.S.)_{\sigma}$. Therefore, the variation of $2\pi T D_{s}^{||,\perp,\times}$ are mostly dictated by the behavior of $\tau_{c}^{||,\perp,\times}(T)$. In a nutshell, the rotation introduces a new time-scale $\tau_{\Omega}$ in the system apart from the usual relaxation scale $\tau_{c}$ and the variation of the conductivity and spatial diffusion components against temperature is determined by the interplay of the $P.S.(T)$ and the ratio of two-time scales--$\tau_{c}(T)$ and $\tau_{\Omega}$.

\begin{figure}
	\centering
	\includegraphics[width=0.45\textwidth]{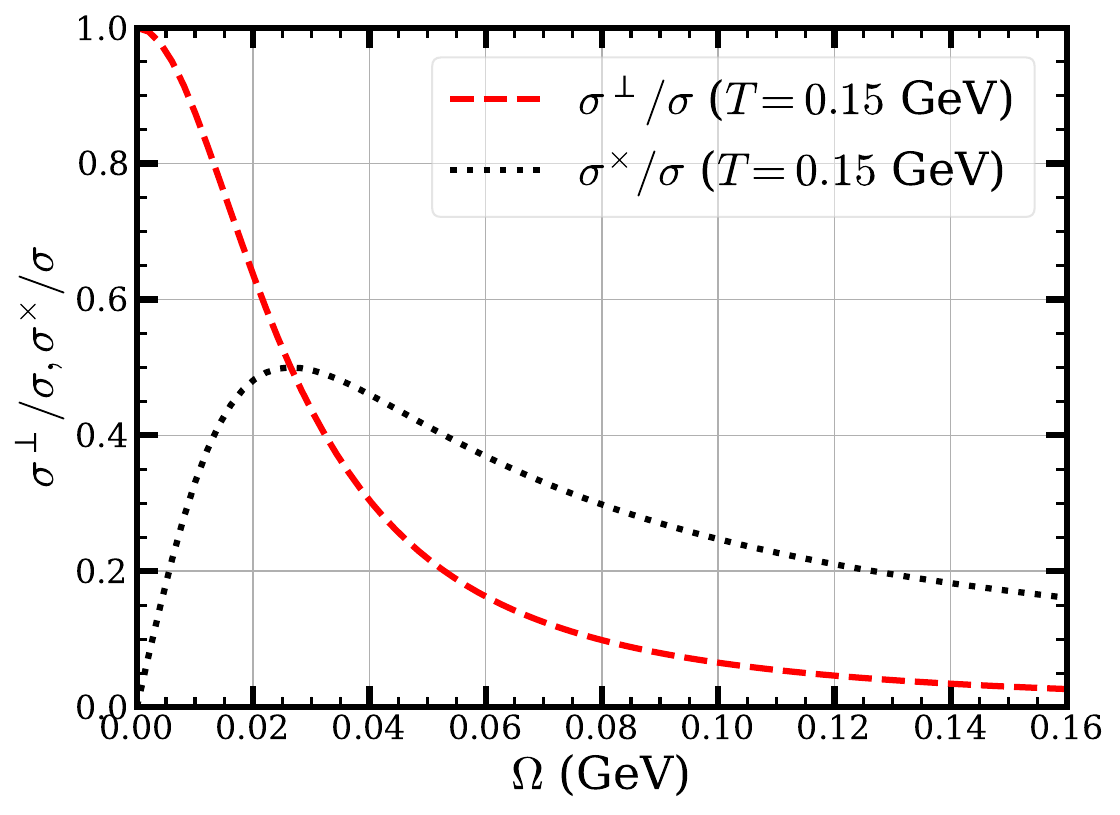}
	\includegraphics[width=0.45\textwidth]{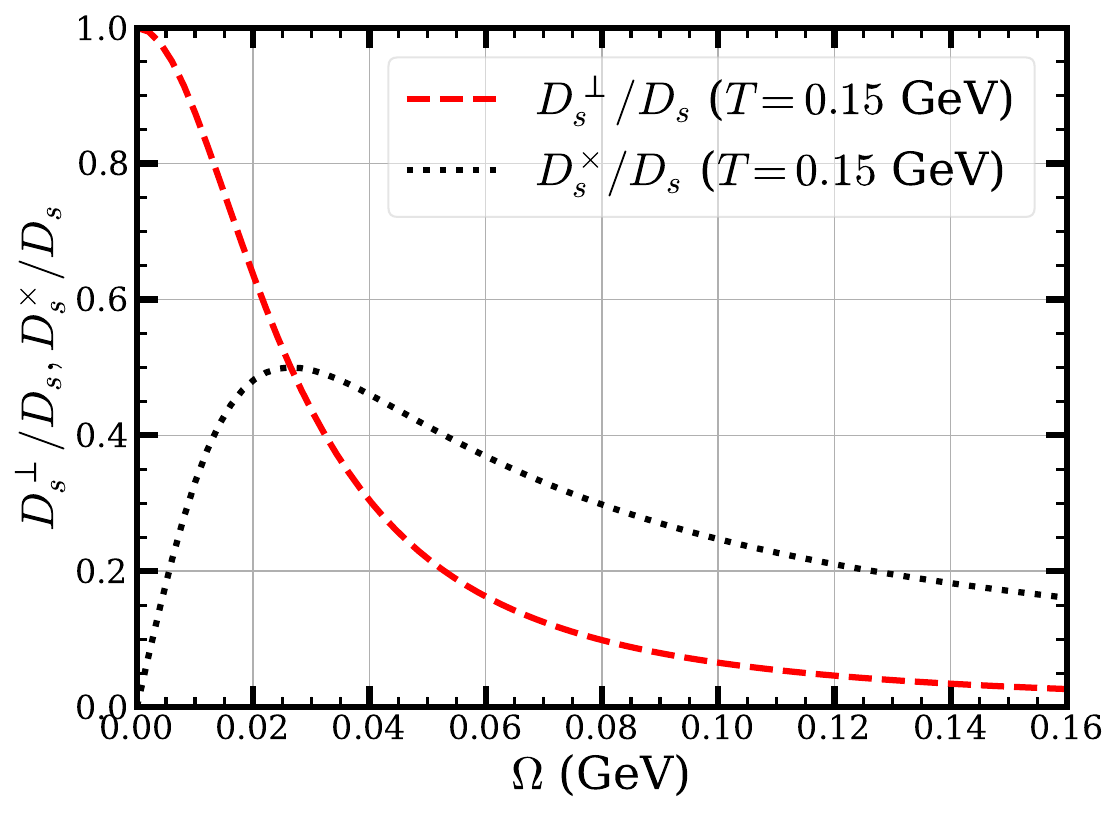}
	\captionsetup{justification=centering, singlelinecheck=off}
	\caption{(Color online) Left: Perpendicular and Hall conductivity of D meson  ($\sigma^{\perp}$ / $\sigma$ , $\sigma^{\times}$ / $\sigma$) as functions of ${\Omega}$. Right:  Perpendicular and Hall spatial diffusion coefficients of D meson ($D^{\perp}_{s}$ / $D_{s}$, $D^{\times}_{s}$ / $D_{s}$) as functions of ${\Omega}$.}
	\label{figure4}
\end{figure}

The left panel of Fig.~(\ref{figure4}) shows the variation of perpendicular and Hall components (normalized) of conductivity as a function of angular velocity $\Omega$. In the absence of rotation, that is, when $\Omega = 0$, $\sigma^{\perp}/\sigma = 1$ and $\sigma^{\times}/\sigma = 0$. The right panel of Fig.~(\ref{figure4}), shows the variation of perpendicular ($D^{\perp}_{s}$) and Hall ($D^{\times}_{s}$) diffusion coefficients (normalized) against angular velocity $\Omega$. As we have taken the ratio of the diffusion components in this case, therefore we are getting a full resemblance with the left panel of the Fig.~(\ref{figure4}). The peak in the Hall conductivity arises because of the structure of effective relaxation time $\tau^{\times}_{c}=\frac{\tau_{c}(\tau_{c}/\tau_{\Omega})}{1+(\tau_{c}/\tau_{\Omega})^{2}}=\frac{2\tau^{2}_{c}\Omega}{1+(2\tau_{c}\Omega)^{2}}$, which has maxima when the two time scales of the system become equal, i.e., when $\tau_{c}=\tau_{\Omega}, \implies \Omega=1/(2\tau_{c})$.
		
In the absence of rotation, the spatial diffusion matrix is diagonal, with the diagonal entries being the same, indicating the same diffusion coefficients in all three spatial directions. However, in the rest frame of the rotating medium, the spatial diffusion matrix is off-diagonal as $D_{s}^{\times}\neq 0$; moreover, the diagonal entries are not the same since $D_{s}^{\perp}\neq D_{s}^{||}$. In the previous Figs.~(\ref{figure3}) and~(\ref{figure4}), we noticed the Hall diffusion coefficient arises as a novel transport coefficient and is significant in the temperature and angular velocity range we have considered. Keeping aside $D_{s}^{\times}$ and following the literature of the systems at finite magnetic fields (in the inertial (laboratory) frame), we may define $D_{s}^{||}-D_{s}^{\perp}$ as the anisotropy induced because of $\vec{\Omega}$ (in the rest frame of the rotating medium). From Figs.~(\ref{figure3}) and (\ref{figure4}), it is evident that at high temperature and low angular velocity, the perpendicular components of conductivity and diffusion approach the parallel components, thereby making the system less anisotropic.

	

After the general discussions on the spatial diffusion coefficients of D meson and its dependence on $\Omega$ and $T$, we now discuss some phenomenological consequences of the D meson diffusion in rotating HRG matter. The nuclear modification factor $R_{AA}$ of D mesons, an essential observable in relativistic HICs, is defined as the ratio of the D meson's final to initial momentum distribution. Our calculations suggest that in the rest frame of the rotating medium anisotropic diffusion (as $D^{||}_{s}  > D^{\perp}_{s}$) and Hall diffusion ($D^{\times}_{s}$) can have a role in the overall modification of D meson distribution function. In the spirit of the framework presented in this paper, one can quantify the effect of a rotating medium on $R_{AA}$ in two steps. Firstly, by solving the transport equation for the distribution function of the D meson in the rest frame of the rotating medium and, in the next step, transforming the D meson distribution to the laboratory frame to get $R_{AA}$, which is measured in the laboratory frame. Obtaining $R_{AA}$ in the proposed method looks promising, and we plan to determine and compare it with the experimental group $R_{AA}$ results shortly.

In this paper, we consider the diffusion of D meson in the rest frame of the rotating medium by introducing pseudo forces like the Coriolis force. So, angular velocity dependency in the diffusion coefficients appears via the Coriolis force. However, in the lab frame, without considering the Coriolis force, the angular velocity dependency can also appear via the interaction of the D meson with the rotating medium constituents. It will be interesting to explore this in the future.

\section{Summary}
\label{Sec:Summary}
In this work, we calculate the diffusion coefficients of D mesons diffusing through the background of a rotating hadron gas. To derive the expressions for these diffusion coefficients, we first formulate a relaxation time-approximated Boltzmann transport equation in a rotating frame. We model the background hadron gas using the hadron resonance gas model and determine the relaxation time for the interactions of D meson with the hadron resonance gas using the hard sphere scattering interactions. We treat the scattering length as a free parameter to adjust the relaxation time. 
To obtain the diffusion coefficients of D meson, we first calculate its conductivity by employing the Boltzmann transport equation. We then relate the diffusion of heavy mesons to their conductivity according to Einstein's diffusion relation, which states that the diffusion coefficients are the ratio of conductivity to susceptibility. The Coriolis force present in the force term of the Boltzmann transport equation is the cause of the anisotropic nature of diffusion of D meson. The tensor structure of both the meson conductivity and diffusion are similar and can be encoded in three components: the parallel, perpendicular, and Hall. The parallel component is independent of angular velocity ($\Omega$) and is equal to the conductivity or diffusion in the absence of rotation. Due to the finite rotation of the medium, parallel and perpendicular diffusion or conductivity components become different. So, anisotropic conductivity or diffusion matrices will be produced at finite rotation via this Coriolis force. Along with the anisotropic structure of diffusion or conduction, a new directional Hall component is induced completely due to the finite rotation of the medium, as it was absent in the non-rotating scenario.
To accurately depict the diffusion coefficients, we first adjusted the relaxation time by tuning the scattering length from $0.18$ fm to $0.85$ fm, aligning with previous estimates of diffusion coefficients for a medium with $\Omega=0$. 
Following this calibration, we examined the variations of normalized parallel, perpendicular, and Hall conductivity (normalized by $T$), as well as diffusion coefficients (multiplied by $2\pi T$), as functions of temperature and angular velocity of the HRG, setting $\tau_{c}$ corresponding to $a=0.85$ fm. We notice non-zero Hall diffusion and conductivity components of the D meson, which is non-monotonic and significant in the temperature and angular velocity values considered. The anisotropy, where parallel and perpendicular diffusion components are different in magnitude, becomes stronger at low temperatures and high angular velocity regimes. So, an isotropic diffusion tensor of D meson and a lower Hall coefficient can be expected at very high temperatures and/or vanishing angular velocity domains.

\section{Acknowledgement}
This work was partially supported by the Ministry of Education, Govt. of India (A.D., N.P., D.R.M.); and the Board of Research in Nuclear Sciences and Department of Atomic Energy, Govt. of India, under Grant No. 57/14/01/2024-BRNS/313 (S.G.). One of the authors (A.D.) thank Prof. Raghunath Sahoo for fruitful discussion.
\appendix
	
	\section{HEAVY MESON CONDUCTIVITY FROM RELATIVISTIC BOLTZMANN EQUATION}
	\label{sec:ECB}
In this appendix, we will provide the derivation of Eq.~(\ref{ancon}) and Eq.~(\ref{ancon1}) with the help of BTE. Substituting $f = f_{0} +\delta f$ we can rewrite Eq.~(\ref{A7}) as follows:
\bea
&&p^{\mu} \frac{\partial f_{0}}{\partial x^{\mu}} - \Gamma_{\mu \lambda}^{\alpha} p^{\mu} p^{\lambda}\frac{\partial f_{0}+\delta f}{\partial p^{\alpha}}= - (u^{\alpha} p_{\alpha}) \frac{\delta f}{\tau_{c}}\nn\\
\implies&&-f
_{0}(1+f_{0})\bigg[\frac{p^{\mu}p^{\alpha}}{T}(\del_{\mu}u_{\alpha}-\Gamma^{\sigma}_{\mu\alpha}u_{\sigma}) + p^{\mu}(u^{\alpha}p_{\alpha})\del_{\mu}\left(\frac{1}{T}\right)-p^{\mu}\del_{\mu}\left(\frac{\mu_{D}}{T}\right)\bigg]\nn\\
&&-\Gamma^{\sigma}_{\mu\lambda}p^{\mu}p^{\lambda}\frac{\del \delta f}{\del p^{\sigma}}=- (u^{\alpha} p_{\alpha}) \frac{\delta f}{\tau_{c}}\nn\\
\implies&&-f_{0}(1+f_{0})\bigg[\frac{p_{0}}{\sqrt{g_{00}}}p^{0}\del_{0}\left(\frac{1}{T}\right)+\frac{p_{0}}{\sqrt{g_{00}}}p^{i}\nabla_{i}\left(\frac{1}{T}\right)-p^{0}\del_{0}\left(\frac{\mu_{D}}{T}\right)-p^{i}\nabla_{i}\left(\frac{\mu_{D}}{T}\right)\bigg]\nn\\
&&+ 2p_{0}(\vec{p}\times \vec{\Omega})\cdot \frac{\del \delta f}{\del \vec{p}}=-\frac{p_{0}}{\sqrt{g_{00}}} \frac{\delta f}{\tau_{c}},\nn\\
\implies&& -f_{0}(1+f_{0})\bigg[\frac{1}{T^{2}}\frac{p^{i}}{E}(\mu_{D}-E)\nabla_{i}T-\frac{p^{i}}{ET}\nabla_{i}\mu_{D}\bigg] + 2(\vec{p}\times \vec{\Omega})\cdot \frac{\del \delta f}{\del \vec{p}}=-\frac{\delta f}{\tau_{c}},  \label{A9}
\eea
where we implicitly assumed that the Greek indices run from $0 \text{ to } 4$ and Latin index $i$ run from $0 \text{ to } 3$; also, we defined $p_{0}\equiv E$. In the simplification process of obtaining Eq.~(\ref{A9}) from Eq.~(\ref{A7}), we used the following approximations: the terms which are 1st order in $\Omega x$, $\Omega y$, and $\frac{\Omega}{T}$ have been retained and the time derivatives of $\mu_{D}$ and $T$ have been ignored assuming a steady state condition~\cite{Padhan:2024edf}. In Eq.~(\ref{A9}) keeping only the thermodynamic force $\nabla_{i}\mu_{D}$, which is responsible for diffusion we have,
\bea
-\frac{\del f_{0}}{\del E} \frac{p^{i}}{E} \nabla_{i}\mu_{D} + 2(\vec{p}\times \vec{\Omega})\cdot \frac{\del \delta f}{\del \vec{p}}=-\frac{\delta f}{\tau_{c}}\label{A10}.
\eea  
For the calculation of current density $J^{i}$, we have to solve Eq.~(\ref{A10}) for $\delta f$. A glance at Eq.~(\ref{A10}) suggest that the solution $\delta f$ should have the following form: $\delta f=-\vec{p}\cdot\vec{X}\frac{\del f_{0}}{\del E}$, where $\vec{X}$ is an arbitrary vector. The vector $\vec{X}$ can be decomposed in terms of the available basis vector at our hand, $\hat{\mu}_{D}=\frac{-\vec{\nabla} \mu_{D}}{|\vec{\nabla} \mu_{D}|}$, $\hat{\omega}\equiv \frac{\vec{\Omega}}{|\vec{\Omega}|}$, and $\hat{\mu}_{D}\times \hat{\omega}$ as: $\vec{X}=\alpha \hat{\mu}_{D}+\beta \hat{\omega}+\gamma (\hat{\mu}_{D}\times \hat{\omega})$ with the unknowns $\alpha$, $\beta$, and $\gamma$. Therefore, the final task boils down to determining the unknowns $\alpha$, $\beta$, and $\gamma$ by substituting $\delta f$ in Eq.~(\ref{A10} )as follows:
\bea
&&\frac{\partial f_0}{\partial E} \frac{\vec{p}}{E} \cdot (-\vec{\nabla} \mu_D) + 2 (\vec{p}\times\vec{\Omega})\cdot \frac{\del}{\del\vec{p}} \left(-\vec{p}\cdot\vec{X}\frac{\del f_{0}}{\del E}\right) = \frac{\vec{p}\cdot\vec{X}}{\tau_{c}}\frac{\del f_{0}}{\del E}\nn\\
\implies&&\frac{\partial f^0}{\partial E} \frac{\vec{p}}{E} \cdot (-\vec{\nabla} \mu_D) -2 (\vec{p}\times \vec{\Omega})\cdot \vec{X}\frac{\del f}{\del E}= \frac{\vec{p}\cdot\vec{X}}{\tau_{c}}\frac{\del f_{0}}{\del E}\nn\\
\implies&&\frac{\del f_{0}}{\del E} \vec{p}\cdot \bigg[-\frac{1}{E}\vec{\nabla}\mu_{D}-2(\vec{\Omega}\times\vec{X})\bigg]=\frac{\vec{X}}{\tau_{c}}\frac{\del f_{0}}{\del E}\cdot \vec{p}\nn\\
\implies && -\frac{1}{E}\vec{\nabla}\mu_{D}-2(\vec{\Omega}\times\vec{X})=\frac{X}{\tau_{c}}.\label{A11}
\eea
Substituting the result $2(\vec{X}\times\vec{\Omega})=(\alpha \hat{\mu}_{D}+\beta \hat{\omega}+\gamma (\hat{\mu}_{D}\times \hat{\omega}))\times \Omega \hat{\omega}=2\alpha\Omega(\hat{\mu}_{D}\times \hat{\omega})-2\gamma\Omega\hat{\mu}_{D}+2\gamma\Omega(\hat{\mu}_{D}\cdot\hat{\omega})\hat{\omega}$, in Eq.~(\ref{A11}) we have,
\bea
\left(\frac{|\vec{\nabla}\mu_{D}|}{E}-\frac{\gamma}{\tau_{\Omega}}\right)\hat{\mu}_{D}+\frac{\gamma}{\tau_{\Omega}}(\hat{\mu}_{D}\cdot\hat{\omega})\hat{\omega}+\frac{\alpha}{\tau_{\Omega}}(\hat{\mu}_{D}\times \hat{\omega})=\frac{\alpha}{\tau_{c}}\hat{\mu}_{D}+ \frac{\beta}{\tau_{c}}\hat{\omega}+\frac{\gamma}{\tau_{c}}(\hat{\mu}_{D}\times\hat{\omega}),\label{A12}
\eea
where we defined $\tau_{\Omega}\equiv \frac{1}{2\Omega}$. Simplifying Eq.~(\ref{A12}), one obtain the following expressions for the unknowns $\alpha$, $\beta$ and $\gamma$,
\begin{align*}
	\alpha &=\frac{|\vec{\nabla}\mu_{D}|}{E} \frac{\tau_c} {1+\big(\frac{\tau_c}{\tau_{\Omega}} \big)^2} , & 
	\beta &= \left(\frac{\tau_c}{\tau_{\Omega}}\right)^2 (\hat{\omega}\cdot\hat{\mu}_{D}) \frac{|\vec{\nabla}\mu_{D}|}{E}\frac{\tau_c}{1+\big(\frac{\tau_c}{\tau_{\Omega}}\big)^2}, &
	\gamma &=\left(\frac{\tau_c}{\tau_{\Omega}}\right) \frac{|\vec{\nabla}\mu_{D}|}{E} \frac{\tau_c} {1+\big(\frac{\tau_c}{\tau_{\Omega}}\big)^2} .
\end{align*}
The $\delta f$ upon substitution of  $\alpha$, $\beta$ and $\gamma$ becomes,
\bea
\delta f&=&-p^{j} X^{j}\frac{\del f_{0}}{\del E}\nn\\
&=& -\frac{\del f_{0}}{\del E}p^{j} (\alpha \hat{\mu}_{D}^{j}+\beta \omega^{j}+\gamma (\hat{\mu}_{D}\times\hat{\omega})^{j})\nn\\
&=&-\frac{\del f_{0}}{\del E}p^{j} (\alpha \hat{\mu}_{D}^{j}+\beta \omega^{j}+\gamma \epsilon^{jkl} \hat{\mu}_{D}^{k}\omega^{l})\nn\\
&=&  -\frac{\del f_{0}}{\del E}\frac{\tau_c} {1+\big(\frac{\tau_c}{\tau_{\Omega}} \big)^2}\frac{p^{j}}{E}\bigg[|\vec{\nabla}\mu_{D}|\hat{\mu}_{D}^{j}+ \left(\frac{\tau_{c}}{\tau_{\Omega}}\right)^{2} \omega^{j}\omega^{k}\hat{\mu}_{D}^{k}|\vec{\nabla}\mu_{D}|+ \frac{\tau_{c}}{\tau_{\Omega}}|\vec{\nabla}\mu_{D}| \epsilon^{jkl}\hat{\mu}_{D}^{k}\omega^{l}\bigg]\nn\\
&=& -\frac{\del f_{0}}{\del E}\frac{p^{j}}{E}\frac{\tau_c}{1+\big(\frac{\tau_c}{\tau_{\Omega}} \big)^2}\bigg[\delta^{ij}+ \frac{\tau_{c}}{\tau_{\Omega}}\epsilon^{jik}\omega^{k}+\left(\frac{\tau_{c}}{\tau_{\Omega}}\right)^{2}\omega^{i}\omega^{j}\bigg](-\nabla_{i}\mu_{D})~.\label{A13}
\eea
The current density $J^{i}$ can now be expressed as,
\bea
J^{i}&=&\int \frac{d^{3}\vec{p}}{(2\pi)^{3}}\frac{p^{i}}{E}\delta f\nn\\
&=&\int \frac{d^{3}\vec{p}}{(2\pi)^{3}}\frac{p^{i}p^{j}}{E^{2}}\left(-\frac{\del f_{0}}{\del E}\right)\frac{\tau_c}{1+\big(\frac{\tau_c}{\tau_{\Omega}} \big)^2}(-\nabla_{k}\mu_{D})\bigg[\delta^{kj}+ \frac{\tau_{c}}{\tau_{\Omega}}\epsilon^{jkl}\omega^{l}+\left(\frac{\tau_{c}}{\tau_{\Omega}}\right)^{2}\omega^{k}\omega^{j}\bigg]\nn\\
&=&\int \frac{d^{3}p}{(2\pi)^{3}}\frac{p^{2}}{3E^{2}}\delta^{ij}\left(-\frac{\del f_{0}}{\del E}\right)\frac{\tau_c}{1+\big(\frac{\tau_c}{\tau_{\Omega}} \big)^2}(-\nabla_{k}\mu_{D})\bigg[\delta^{kj}+ \frac{\tau_{c}}{\tau_{\Omega}}\epsilon^{jkl}\omega^{l}+\left(\frac{\tau_{c}}{\tau_{\Omega}}\right)^{2}\omega^{k}\omega^{j}\bigg], (d^{3}p\equiv 4\pi p^{2}dp)\nn\\
&=&\int \frac{d^{3}p}{(2\pi)^{3}}\frac{p^{2}}{3E^{2}}\left(-\frac{\del f_{0}}{\del E}\right)\frac{\tau_c}{1+\big(\frac{\tau_c}{\tau_{\Omega}} \big)^2}(-\nabla_{k}\mu_{D})\bigg[\delta^{ki}+ \frac{\tau_{c}}{\tau_{\Omega}}\epsilon^{ikl}\omega^{l}+\left(\frac{\tau_{c}}{\tau_{\Omega}}\right)^{2}\omega^{k}\omega^{i}\bigg]\nn\\
&=&\frac{1}{T}\int \frac{d^{3}p}{(2\pi)^{3}}\frac{p^{2}}{3E^{2}}\frac{\tau_c}{1+\big(\frac{\tau_c}{\tau_{\Omega}} \big)^2}(-\nabla_{j}\mu_{D})\bigg[\delta^{ij}+ \frac{\tau_{c}}{\tau_{\Omega}}\epsilon^{ijk}\omega^{k}+\left(\frac{\tau_{c}}{\tau_{\Omega}}\right)^{2}\omega^{i}\omega^{j}\bigg]f_{0}(1+f_{0})~.\label{A14}
\eea
Comparing it with Eq.~(\ref{macon}) we obtain the conductivity matrix as follows:
\bea
\sigma^{ij}=\sigma^{0}\delta^{ij}+\sigma^{1}\epsilon^{ijk}\omega^{k}+\sigma^{2}\omega^{i}\omega^{j}\label{A15},
\eea
where $\sigma^{n}$ are expressed as,
\bea
&& \sigma^{n}=\frac{1}{T}\int \frac{d^{3}p}{(2\pi)^{3}}\frac{p^{2}}{3E^{2}}f_{0}(1+f_{0})\frac{\tau_{c}(\tau_{c}/\tau_{\Omega})^{n}}{1+(\tau_{c}/\tau_{\Omega})^{2}}~.\label{A15}
\eea

\bibliographystyle{unsrturl}
\bibliography{Dmesonref}
\end{document}